\documentclass[sigconf]{acmart}

\usepackage{multirow}
\usepackage{amsmath}

\usepackage{algorithm}
\usepackage{algpseudocode}

\usepackage[detect-all]{siunitx}

\usepackage{xcolor}

\AtBeginDocument{%
  \providecommand\BibTeX{{%
    \normalfont B\kern-0.5em{\scshape i\kern-0.25em b}\kern-0.8em\TeX}}}

\copyrightyear{2021}
\acmYear{2021}
\setcopyright{rightsretained}
\acmConference[HPDC '21]{Proceedings of the 30th International Symposium on High-Performance Parallel and Distributed Computing}{June 21--25, 2021}{Virtual Event, Sweden}
\acmBooktitle{Proceedings of the 30th International Symposium on High-Performance Parallel and Distributed Computing (HPDC '21), June 21--25, 2021, Virtual Event, Sweden}\acmDOI{10.1145/3431379.3460650}
\acmISBN{978-1-4503-8217-5/21/06}

\begin{document}

\fancyhead{}

%%
%% The "title" command has an optional parameter,
%% allowing the author to define a "short title" to be used in page headers.
\title{Q-adaptive: A Multi-Agent Reinforcement Learning Based Routing on Dragonfly Network}

%%
%% The "author" command and its associated commands are used to define
%% the authors and their affiliations.
%% Of note is the shared affiliation of the first two authors, and the
%% "authornote" and "authornotemark" commands
%% used to denote shared contribution to the research.

\author{Yao Kang}
\email{ykang17@hawk.iit.edu}
\affiliation{%
  \institution{Illinois Institute of Technology}
  \city{Chicago}
  \state{Illinois}
  \country{USA}
}

\author{Xin Wang}
\email{xwang149@hawk.iit.edu}
\affiliation{%
  \institution{Illinois Institute of Technology}
  \city{Chicago}
  \state{Illinois}
  \country{USA}
}

\author{Zhiling Lan}\thanks{Zhiling Lan's current affiliation is University of Illinois Chicago, and her current contact is \ttfamily\upshape{zlan@uic.edu}.}
\email{lan@iit.edu}
\affiliation{%
  \institution{Illinois Institute of Technology}
  \city{Chicago}
  \state{Illinois}
  \country{USA}
}

%%
%% By default, the full list of authors will be used in the page
%% headers. Often, this list is too long, and will overlap
%% other information printed in the page headers. This command allows
%% the author to define a more concise list
%% of authors' names for this purpose.

% \renewcommand{\shortauthors}{XX and YY, et al.}

%%
%% The abstract is a short summary of the work to be presented in the
%% article.
\begin{abstract}

High-radix interconnects such as Dragonfly and its variants rely on adaptive routing to balance network traffic for optimum performance.
Ideally, adaptive routing attempts to forward packets between minimal and non-minimal paths with the least congestion.
In practice, current adaptive routing algorithms estimate routing path congestion based on local information such as output queue occupancy.
Using local information to estimate global path congestion is inevitably inaccurate because a router has no precise knowledge of link states a few hops away.
This inaccuracy could lead to interconnect congestion.
In this study, we present Q-adaptive routing, a multi-agent reinforcement learning routing scheme for Dragonfly systems.
Q-adaptive routing enables routers to learn to route autonomously by leveraging advanced reinforcement learning technology.
The proposed Q-adaptive routing is highly scalable thanks to its fully distributed nature without using any shared information between routers.
Furthermore, a new two-level Q-table is designed for Q-adaptive to make it computational lightly and saves 50\% of router memory usage compared with the previous Q-routing.
We implement the proposed Q-adaptive routing in SST/Merlin simulator.
Our evaluation results show that Q-adaptive routing achieves up to 10.5\% system throughput improvement and 5.2x average packet latency reduction compared with adaptive routing algorithms.
Remarkably, Q-adaptive can even outperform the optimal VALn non-minimal routing under the ADV+1 adversarial traffic pattern with up to 3\% system throughput improvement and 75\% average packet latency reduction.

\end{abstract}

%%
%% The code below is generated by the tool at http://dl.acm.org/ccs.cfm.
%% Please copy and paste the code instead of the example below.
%%



%%
%% Keywords. The author(s) should pick words that accurately describe
%% the work being presented. Separate the keywords with commas.

%%
%% This command processes the author and affiliation and title
%% information and builds the first part of the formatted document.
\maketitle

\section{Introduction}

Interconnect network is a critical component in high-performance computing (HPC) systems. 
It serves as a “central nervous system” for data exchange between computer nodes \cite{DCnetworking}.
Recent high-radix interconnect such as Dragonfly \citep{08df} and its variants \citep{sc12cray}\citep{df+} become a dominant interconnect topology in Top500 supercomputers \cite{top500}.
The Slingshot interconnects \citep{sc20slingshot}, which will power future exascale HPC systems, also adopt Dragonfly topology for its cost-effective, low-diameter, and highly scalable features.

Dragonfly networks have a hierarchical design, in which system nodes are divided into several fully connected, identical groups.  
A dragonfly system is highly scalable due to the use of high-radix routers and a fully connected group structure.
New groups can be flexibly added to the system while maintaining its low-diameter quality. 
With the fully connected groups, Dragonfly provides \textit{high path diversity} such that packets can be forwarded either minimally between source and destination groups or non-minimally through any intermediate group in the system.
High path diversity allows routing to take diverse paths under different network patterns: 
\textit{minimal routing} is optimal for balanced distributed traffic such as uniform random traffic pattern, 
whereas \textit{Valiant non-minimal routing (VAL routing)} is advantageous in case of unbalanced traffic such as adversarial traffic pattern.

Adaptive routing is often deployed in Dragonfly systems by dynamically delivering packets either minimally or non-minimally according to real-time network conditions \cite{08df}\cite{isca09indirectadp}. 
Universal Globally-Adaptive Load-balanced routing (UGAL) is widely used in production dragonfly systems \cite{sc12cray}\cite{sc20slingshot}.
It allows the source router to choose between minimal or non-minimal routing paths with the least estimated congestion.
Progressive Adaptive Routing (PAR) enhances UGAL by enabling source group routers to re-evaluate previous routing decisions based on the current estimated congestion \cite{isca09indirectadp}.
Because no system-wide global information is shared among routers, both UGAL and PAR rely on local information such as output queue occupancy to estimate routing path congestion. 
As a switched fabric network, source and destination routers are often connected through several hops.
Thus, local information can only be a good indicator for near-end congestion and often fails to infer far-end congestion \cite{dffarendcongest} or downstream congestion \cite{isca09indirectadp} at neighbor and downstream hops.
\begin{figure}[ht]
  \centering
  \includegraphics[scale=0.9]{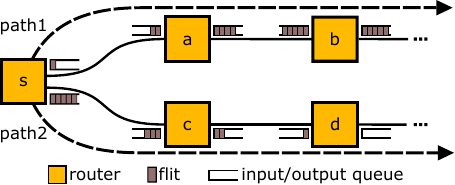}
  \caption{Issue of existing adaptive routing on Dragonfly, where local information fails to estimate global path congestion.
  Although path2 is the least congested path, existing adaptive routing methods typically choose path1 over path2.}
  \Description{}
  \label{fig:linkcongest}
\end{figure}
A simple example is illustrated in Figure \ref{fig:linkcongest} to show the limitation of UGAL and PAR. 
In this example, router $s$ prefers $path1$ over $path2$ because the former has a less occupied output queue at router $s$.
However, selecting $path1$ in this scenario is inappropriate because of the congestion at downstream routers of $a$ and $b$.
This illustrative example clearly demonstrates that adaptive routing based on local information (e.g., local output queue occupancy) could cause wrong decision making, hence leading to network congestion, delayed packet delivery, and limited system throughput.

In this work, we present \textit{a learning-based routing method to autonomously learn network conditions on Dragonfly systems by leveraging advanced machine learning technology.} 
Specifically, we choose reinforcement learning (RL) over supervised learning as the latter requires human effort to provide true labels, or desirable routing behaviors during the training phase, which can be practically impossible \cite{17hotnetLearntoroute}. 
Unlike supervised learning, RL agents learn to behave by interacting with an environment for an objective, which can be mathematically expressed to maximize a cumulative reward.
RL has been successfully applied for car driving system \cite{15selfdriving}, human-level gaming \cite{dqn}, etc.
Recent studies have shown great promise of reinforcement learning in successfully solving computer system problems such as job scheduling \cite{ipdps21yuping}\cite{19schedulingmao}, memory management \cite{rlMemManage}, circuit design \cite{rlCircuitDesign}, NoC arbitration \cite{hpca20nocarbitration}, etc.

A pioneering work in applying reinforcement learning to network routing was done by Boyan et al. in 1993 \cite{93qrting}.  
They proposed Q-routing, a multi-agent reinforcement learning (MARL) method that each router works as an independent agent and is guided by a state-action lookup table, Q-table, for routing decisions. 
In their study, Q-routing was evaluated on an irregular $6 \times 6$ grid network.
Later, several enhancements were presented to improve the quality of Q-routing \cite{96pqrting}\cite{98bcastq}.
These studies mainly focused on similar irregular grid networks at a small scale (e.g., 15--36 routers). 
% \textcolor{blue}{For large-scale systems, MARL routing approach is preferred over single-agent RL solutions because the central RL agent requires additional wire connections to the routers making routing design complex and limiting system size.} \textcolor{red}{(I suggest to delete this sentence because there is no need to mention central RL.)}
% However, applying Q-routing or designing new MARL routing algorithms for modern high-radix Dragonfly networks is challenging.

\emph{Computer systems have been evolved dramatically over the past twenty years in terms of interconnect topology as well as the system scale.}
Little is known about the feasibility and efficiency of applying MARL for routing in a large-scale Dragonfly system. 
Previous MARL studies, e.g., Q-routing, were focused on grid networks \cite{93qrting}\cite{96pqrting}\citep{98bcastq}, which are significantly different from the hierarchical, high-radix Dragonfly networks.
Furthermore, modern supercomputers typically contain hundreds or thousands of routers. 
Practically training a RL model with a large number of independent agents to a converged state is much more challenging than the previous small-scale case studies. 
HPC network traffic is constantly changing with different patterns and loads. 
This demands the training to converge in a reasonable time under both uniform and adversarial traffic patterns whereas the previous studies only considered the uniform random traffic pattern \cite{93qrting}\citep{98bcastq}\cite{01pgrouting}\cite{02pgrouting}.
Moreover, considering computational resources on routers are limited, we must avoid any RL methods that have a high computational requirement, e.g., deep learning methods using neural networks \cite{19dlrting}\cite{20dlrting}.

To tackle the above challenges, we present \textit{Q-adaptive routing}, a fully distributed MARL routing scheme for Dragonfly networks.
Inspired by the Q-routing studies, Q-adaptive routing adopts table-based RL for decision making. 
Table-based RL methods are computationally efficient as compared to deep neural network based methods \cite{19dlrting}\cite{20dlrting}, hence being more realistic for practical use.  
Distinguishing from Q-routing, Q-adaptive utilizes a novel two-level Q-table. 
The two-level Q-table not only provides more learning information, but also mitigates outdated Q-value issue commonly experienced in large-scale systems due to spare updates.
Moreover, our two-level Q-table only requires half of the memory usage of the Q-table used in Q-routing. 
New techniques are adopted for Q-adaptive routing to ensure timely update of values in the two-level Q-table, hence guarantee a fast and stable model convergence. 
Finally, Q-adaptive assures packets to be delivered within five hops, which guarantees a routing livelock and deadlock free design for Dragonfly networks (details in Section \ref{sec:qadp}).

% Q-adaptive is a highly scalable routing algorithm without requiring any shared information between agents compared with other approaches \cite{01pgrouting}\cite{02pgrouting}.
% \textcolor{blue}{
% Q-adaptive routing extends Q-routing in several aspects. 
% First, Q-adaptive routing adopts a two-level Q-table, which only requires half of the memory usage of the original Q-table used in Q-routing.
% Table-based RL routing method is computational efficiently as compared with deep neural network based methods 

We implement our design in SST/Merlin, a flit-level event-driven simulation toolkit \cite{sst}. 
We evaluate Q-adaptive routing on a 1,056-node Dragonfly system consisting of 264 routers and further demonstrate its scalability on a 2,550-node system with 510 routers.  
Extensive experiments are conducted to evaluate Q-adaptive routing with several routing methods presented for Dragonfly systems under various traffic patterns and loads. 
Our results show that Q-adaptive routing outperforms the existing adaptive routing methods by up to 10.5\% in network throughput and over five times of reduction in packet latency. 
Moreover, Q-adaptive routing outperforms the optimal routing (e.g., non-minimal routing) under adversarial traffic patterns. 
For system convergence, our results show that Q-adaptive is guaranteed to converge under various traffic patterns and network loads within \SI{500}{\micro\second}.

For the rest of the paper, Section \ref{sec:bckground} presents background and related work.
% Section \ref{sec:methodology} describes our methodology. 
Section \ref{sec:challenge} discusses the challenges of designing MARL based routing for Dragonfly systems.
Section \ref{sec:qadp} gives the details of our work, followed by evaluations in Section \ref{sec:1k-res} and Section \ref{sec:2k-res}. 
Section \ref{sec:conclude} concludes the paper.

\section{Background and related work} \label{sec:bckground}

% In this section, we give an overview of Dragonfly topology, along with existing routing studies. 
% We also introduce reinforcement learning with the discussion of Q-learning and its application in a multi-agent environment.
% In the end, we finalize this section with related work.

\subsection{Dragonfly Topology}

\begin{figure}[ht]
  \centering
  \includegraphics[scale=0.9]{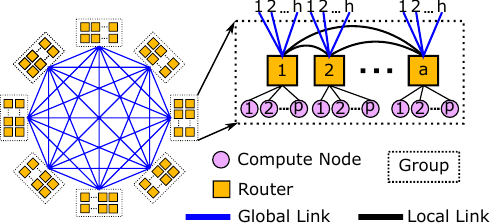}
  \caption{Dragonfly Topology}
  \Description{}
  \label{fig:1ddf}
\end{figure}

Dragonfly \cite{08df} is a high-radix interconnect topology featuring high-bandwidth, low-diameter, and extreme scalability at a reasonable cost.
As shown in Figure \ref{fig:1ddf}, Dragonfly arranges network resources (i.e., routers and links) into multiple identical groups such that the network topology is organized in a three-tiered hierarchy: nodes connections, intra-group connection through local links, and inter-group connection through global links.
Depending on the connected link type, router ports can be classified into host port, local port, or global port for different connection levels in the topology hierarchy.
Dragonfly uses all-to-all inter-group connection such that packets generated in the source group can be minimally forwarded to any destination groups using only one global link hop.

In contrast to a fully connected inter-group connection, the intra-group connection is not limited to a single form.
For example, Figure \ref{fig:1ddf} shows a 1-dimensional fully connected intra-group network as presented in \cite{08df}.
Additionally, the intra-group connection can also be configured with a 2-dimensional partially all-to-all network in Cray Cascade systems \cite{sc12cray}, or a bipartite-graph tree network used in Megafly \cite{megafly} and Dragonfly+ \cite{df+}. 
In this work, we focus on the Dragonfly topology with all-to-all inter-/intra-group connection as shown in Figure \ref{fig:1ddf}, because this structure will be deployed on the Slingshot interconnects to support the next-generation exascale-computing systems \cite{sc20slingshot}. 
We shall point out that the proposed Q-adaptive routing is not limited to a single form and can also be applied to other Dragonfly variants.
In the rest of the paper, we simply refer to this all-level fully connected Dragonfly as Dragonfly and use the nomenclature in Table \ref{tab:1056DF}.

\begin{table}[ht]
  \caption{Dragonfly configurations}
  \label{tab:1056DF}
  \begin{tabular}{c|c|cc}
    \toprule
    Parameter &Note & \multicolumn{2}{c}{Systems}\\
    \midrule
    $N$ & Number of nodes & \multicolumn{1}{c|}{1056} & 2550\\
	$p$ & compute nodes per router & \multicolumn{1}{c|}{4} & 5 \\ 
	$a$ & routers per group & \multicolumn{1}{c|}{8} & 10 \\
	$h$ & global links per router & \multicolumn{1}{c|}{4} & 5 \\
	$k = p+h+a-1$ & ports per router & \multicolumn{1}{c|}{15} & 19 \\
	$g = ah+1$ & number of groups & \multicolumn{1}{c|}{33} & 51 \\
	$m = g*a$ & routers in the system & \multicolumn{1}{c|}{264} & 510 \\
    \bottomrule
  \end{tabular}
\end{table}

With the hierarchical design, Dragonfly is a diameter-3 topology that packets can be minimally delivered within three hops: one local link in the source group, one global link crossing groups, and one local link in the destination group.
Because a minimal path uses twice as much local link bandwidth as that of global link, a proper Dragonfly should be configured with $a = 2p = 2h$ for load balancing purposes \citep{08df}.

In production, Cray's Aries interconnects adopt \emph{credit-based flow control}, which guarantees a packet lossless Dragonfly system.
In such a system, each router maintains a counter, named credits, to record the number of free buffer slots in its downstream routers. 
A flit can only be forwarded to the next router when this number is positive. 
Whenever a flit is forwarded and leaves a router, the router will send back a credit to its upstream router indicating that a new free buffer slot becomes available.

\subsection{Dragonfly Routing} \label{sec:dfrting}

Dragonfly routing is commonly evaluated under the best and the worst case traffic patterns \cite{08df}\cite{isca09indirectadp}:

\textbf{Uniform Random (UR):}
Each node communicates with a randomly selected node on a per message basis.
UR traffic pattern balances network traffic through the randomly chosen source-destination pair, thus is considered as the best case on Dragonfly topology.
Minimal routing typically provides the optimal performance and is expected to achieve 100\% system throughput.

\begin{figure}[ht]
  \centering
  \includegraphics[scale=0.9]{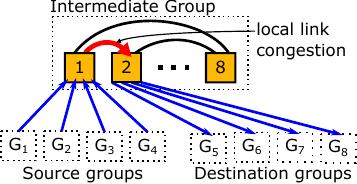}
  \caption{Local link congestion under ADV+4. In this case, G1 send packets to G5, G2 to G6, etc. When the packets are routed non-minimally, the local link between router 1 and router 2 in the intermediate group becomes bottleneck (red arrow).}
  \Description{}
  \label{fig:adv+i}
\end{figure}

\textbf{Adversarial (ADV+i):}
Each node in group $G$ sends packets to a random node in group $G+i$.
This is the worst case as network traffic is extremely unbalanced such that packets generated in one group all target another group.
As a result, the limited number of global links between groups becomes the bottleneck. 
Non-minimal routing via intermediate groups is typically suitable for this traffic pattern and is expected to achieve up to 50\% system throughput. 
Note that ADV+i may also lead to local link congestion in intermediate groups \cite{icpp12adv+i}.
Figure \ref{fig:adv+i} illustrates the potential local link congestion under ADV+4 traffic.
For the 1,056-node Dragonfly system described in Table 1, the ADV+4 traffic pattern has the most local link congestion, whereas ADV+1 has the least congestion.  

%\begin{table}[ht]
%  \caption{Routing on Dragonfly}
%  \label{tab:adprting}
%  \begin{tabular}{c|c|c}
%    \toprule
%    Routing & Note & \#VCs\\
%    \midrule
%    $MIN$  & Minimal routing & 2 \\
%    $VALg$ & Valiant routing through intermediate group & 3\\
%    $VALn$ & Valiant routing through intermediate node & 4\\ 
%	$UGALg$ & minimal path or VALg path & 3 \\ 
%	$UGALn$ & minimal path or VALn path & 4\\
%	$PAR$ & minimal path and VALn path & 5\\
%	$Q-adp$ & Q-adaptive routing & 5\\
%    \bottomrule
%  \end{tabular}
%\end{table}

Dragonfly routing mechanisms can be broadly classified as non-adaptive and adaptive methods. 
\emph{Non-adaptive routing} includes minimal and Valiant routing:

\textbf{Minimal routing (MIN):} 
Packets are minimally forwarded to their destinations within three hops.
Minimal routing is the optimal routing policy for the UR traffic pattern, but the worst choice for the ADV+i traffic pattern.
Minimal routing requires two virtual channels (VCs) to 
avoids routing deadlock \cite{vc}\cite{dally1988deadlock}.

\textbf{Valiant routing (VALg, VALn):}
Valiant non-minimal routing (VAL) is an optimal solution for the ADV+i traffic pattern.
Valiant-global (\textbf{VALg}) routes packets minimally to a random intermediate \emph{group}, then minimally to the destination group \cite{08df}.
A VALg non-minimal path contains up to five hops and requires three VCs to avoid deadlock.
Valiant-node (\textbf{VALn}) solves the local link congestion problem by rerouting packets to reach a specific random \emph{router} in an intermediate group \citep{dffarendcongest}.
As a result, packets are not minimally forwarded in intermediate groups at the cost of an additional local link making VALn uses up to six hops and requires four VCs to avoid deadlock.

Since both MIN and VAL are only good for one traffic pattern,
\emph{adaptive routing} is typically adopted in production systems to let routers dynamically choose between minimal and non-minimal paths \cite{08df}\cite{isca09indirectadp}.
The dynamic decisions depend on network conditions deduced from a router's local queue occupancy.
If the local queue occupancy of a candidate minimal path is less than twice of a candidate non-minimal path, the router will forward the packet minimally.
In practice, a bias can be added to give preference for minimal paths.
% \begin{equation}
% \label{eq:adprouting}
% q_{min} < 2 \times q_{non\_min} + T
% \end{equation}
There are two widely used adaptive routing methods:

\textbf{UGALg, UGALn:} Universal Globally-Adaptive Load-balanced routing (UGAL) \cite{08df} lets source routers make adaptive routing decisions.
\textbf{UGALg} chooses between a minimal path and a VALg non-minimal path, whereas \textbf{UGALn} considers the VALn non-minimal path.

\textbf{PAR:} Progressive Adaptive Routing (PAR) chooses between minimal paths and VALn non-minimal paths \cite{isca09indirectadp}\cite{dffarendcongest}.
In contrast to UGAL, PAR allows source group routers to re-evaluate routing decisions if a packet is being minimally routed.
PAR can achieve better performance than UGAL because it enables source group routers to adjust routing strategies depending on the local congestion, which can be difficult to be perceived by the source router.
However, switching from a minimal path to a non-minimal path costs an additional local link making PAR use up to seven hops.
To avoid routing deadlock, PAR requires five VCs.

There are several other routing related studies. 
Jiang et al. introduced different indirect adaptive routing algorithms to improve the original UGAL \cite{isca09indirectadp}.
Garcia et al. proposed the OFAR routing to mitigate local link congestion and
Won et al. proposed window-based PAR routing to overcome far-end congestion \cite{icpp12adv+i}\cite{dffarendcongest}.
Rahman et al. customized UGAL by limiting the set of candidate non-minimal paths according to Dragonfly configuration \cite{sc19tugal}. 
Chunduri et al. studied run to run performance variability on a Dragonfly production system \cite{sc17run2run}, and simulation studies analyzed network interference with different intra-group connections  \cite{sc16yang}\cite{pads19kang}\cite{ipdps20xin}.
Mubarak et al. and Wilke et al. proposed QoS based approaches to mitigate network interference \cite{misbah19qos}\cite{cluster20qos}. 
Jain et al. proposed to reduce network hot-spots through random job placement \cite{sc14jain}.
De Sensi et al. proposed to reduce network noise using dynamic minimal routing bias \cite{sc19adpthreshold}.
Fundamentally, existing Dragonfly routing methods are based on heuristics by using local queue occupancy for decision making. 
This work leverages multi-agent reinforcement learning (MARL) to train routers to learn beyond queue occupancy for routing. 
% Connors et al. and McGlohon et al. explored link failure effect on Dragonfly class networks \cite{ipdpsw19linkfail}\cite{pads20linkfail}.

\subsection{Reinforcement Learning based Routing}

Reinforcement learning is a subarea of machine learning that automatically learns to maximize cumulative reward through interaction with the environment \cite{rltextbook}.
The environment can be formulated as a Markov Decision Process (MDP), where an agent observes the current environment state, selects an action, observes the environment move to a new state, and receives a feedback reward.
The whole process iterates until a terminal state.
The agent is expected to find an optimal policy that maximizes the cumulative reward.
\emph{Q-learning} is a model-free RL algorithm with the expected cumulative reward named Q-value and uses $\epsilon$-greedy policy to balance the exploration-exploitation dilemma \cite{rltextbook}.
In practice, Q-values can be either stored in Q-table or be approximated by a deep neural network \cite{dqn}.
%\begin{equation}
%\label{eq:qlearning}
%Q(s, a) = Q(s, a) + \alpha * (r + \gamma \cdot \max_{a'} Q(s',a') - Q(s, a)) 
%\end{equation}
When there is more than one agent in the system, it falls into the domain of \emph{multi-agent reinforcement learning (MARL)}.
Depending on the objective, MARL can be either competitive or cooperative \cite{98dynamicsmarl}.
% In addition, Claus et al. \cite{98dynamicsmarl} distinguish two forms of agents in MARL: independent learners (ILs), that agents are only aware of their actions; and joint action learners (JALs), that an agent's action is chosen in conjunction with others based on some shared information in the system.
Routing on large-scale interconnect networks is by nature a MARL problem because routers in the system can be considered as independent agents. In this work, we consider our Dragonfly routing problem as \emph{a cooperative independent MARL process} that routers behave as independent learners for a common objective, that is, to deliver messages in the shortest time. 

\subsubsection{\textbf{Q-routing}}

Boyan et al. proposed \emph{Q-routing}, a MARL routing method based on Q-learning \cite{93qrting}. 
In Q-routing, each router maintains a Q-table of size $m \times (k-p)$, where $m = g \times a$ is the total number of routers in the system, $k$ is the router radix and $p$ is the number of host ports on each router. 
Table \ref{tab:qtable} illustrates a Q-table where each Q-value is the estimated packet delivery time from the current router to a destination router through a corresponding port.
For example, when router $R_x$ receives a packet whose destination is router $R_d$,
$R_x$ will read row $d$ of the Q-table to send the packet through the port with the minimum estimated delivery time i.e., the smallest Q-value.
Q-routing explores the solution space with $\epsilon$-greedy policy, such that a random port with $\epsilon$ probability may be selected as the final outbound port instead of the best port.
All downstream routers forward the packet in the same way.

% Table \ref{tab:qtable} illustrates a Q-table where Q-values are estimated packet delivery time from the current router to different destinations (i.e., routers) through different paths (i.e., ports). 
% Each router maintains its own Q-table. 
% For example, if Table \ref{tab:qtable} is held by a router $R_x$, then a packet for destination router $R_1$ is expected to be delivered using 320ns through port1, which is 20ns longer than using port2.
% Because a router can only forward packets in the network to its neighbor routers through local or global ports, host ports are not included in the table.

\begin{table}[ht]
\caption{A Q-table example, where port 2 will be selected for routing to destination $R_1$.}
\label{tab:qtable}
\begin{tabular}{|c|cccc|}
\hline
\multirow{2}{*}{To Dest.} & \multicolumn{4}{c|}{By Port} \\
                          & $1$     & $2$     & ...   & $k-p$  \\ \hline
$R_1$                        & 320ns   & 300ns   & ...  & 480ns     \\
$R_2$                        & 260ns   & 310ns   & ... & 310ns    \\
...                       &       &      &       &      \\
$R_m (R_{g \times a})$                        & 330ns   & 330ns   & ... & 330ns    \\ \hline
\end{tabular}
\end{table}

As a packet is passed to its destination, routers on the routing path update their Q-tables.
Assuming the previous packet destined for $R_d$ is forwarded from $R_x$ to $R_y$. 
After $R_y$ selects next hop, its smallest Q-value $Q_y$ and a reward $r$ will be sent back to $R_x$.
In Q-routing, the immediate reward is defined as the packet traveling time from $R_x$ to $R_y$.
After receiving the feedback, $R_x$ updates its Q-value $Q_x$ for destination $R_d$ with the Q-learning algorithm shown in Equation (\ref{eq:qrting}), where $\alpha$ is the learning rate. 
\begin{equation} \label{eq:qrting}
\begin{gathered}
\Delta Qx = \alpha (r + Q_y - Q_x) \\
x,y \in N
\end{gathered}
\end{equation}
In Q-routing, routers are independent agents because routing decisions are purely made based on the individual Q-table and there is no explicit shared information between routers.

% Q-routing can lead to outdated Q-values when a destination is rarely visited.
Nowe et al. enhanced Q-routing with periodic Q-value broadcasting to improve the precision of Q-table at the expense of extra network traffic \cite{98bcastq}. 
Tao et al. and Peshkin et al. proposed to use policy gradient and gradient ascent method for MARL routing \cite{01pgrouting}\cite{02pgrouting}. 
However, both require a system-wide global reward, which causes extra network traffic and is impractical for large systems.
Prashanth et al. theoretically applied an actor-critic algorithm for routing in a centralized manner \cite{16AC}.
Deep learning routing approaches were presented in several studies \cite{19dlrting}\cite{20dlrting}; however deep neural networks require heavy computation resources, which makes them unrealistic for practical usage.  

None of the above studies target Dragonfly topology, hence little is known about the feasibility and the performance of MARL routing in dragonfly systems, in particular those deployed with hundreds of routers. 

\subsubsection{\textbf{Issues of Q-Routing on Dragonfly}} \label{sec:qrtingissue}

While Q-routing is a good starting point, it was presented in 1993 and evaluated on a $6 \times 6$ irregular grid network, which is very different from modern Dragonfly systems.
The original Q-routing method does not limit routing path length, which prevents it from being applicable on Dragonfly topology for two reasons:
(1) \emph{routing livelock}, a packet travels in the network without advancing towards its destination;
(2) \emph{routing deadlock}, Dragonfly relies on distributing packets on different virtual channels (VCs) to avoid deadlock.
When a packet traverses in the network, it increases its VC index to break the channel dependency loop \cite{dally1988deadlock}.
Since the number of VCs is limited by the hardware, a packet must reach its destination before exhausting all available VCs.

To address the above issues, a naive approach would be the use of a hop count threshold denoted as \textit{maxQ}. 
Once a packet has visited \textit{maxQ} routers, it will be routed minimally to the destination.
Since Dragonfly is a diameter-3 topology, packets are assured to be delivered within $maxQ + 3$ hops such that livelock can never happen and deadlock can be resolved with a limited number of VCs.

We analyze this naive method by using SST/Merlin simulation \cite{sst}.
Our results indicate that there is no one-size-fits-all value for \textit{maxQ}. 
While UR prefers small \textit{maxQ} to mimic a minimal routing, ADV+i favors larger \textit{maxQ} values to bypass the highly utilized global links between source and destination groups. 
Moreover, extremely large \textit{maxQ} can be detrimental such that packets spend a long time in the network visiting unnecessary routers, resulting in both large packet latency and low network throughput.
Most importantly, regardless of the value of \textit{maxQ}, performance under the ADV+4 pattern is always worse than the ADV+1 pattern.
This indicates that Q-routing is not capable of handling local link congestion.
As a result, it is difficult, if not impossible, to set an appropriate \textit{maxQ} for both UR and ADV+i traffic patterns.
Additionally, since the size of the Q-table is linear to the system size, in the case of large systems, Q-values could be outdated when a destination is rarely visited. 
As such, we observed instability and slow convergence of the enhanced Q-routing method in our experiments. 
In summary, the original Q-routing and the naive Q-routing enhancement do not work for Dragonfly networks.

\section{Technical Challenges} \label{sec:challenge}

Exploring MARL routing for Dragonfly systems poses several key challenges:

\textbf{Topology uniqueness.}
Dragonfly is a hierarchical topology with all-to-all connected router groups.
As a result, the topology provides a large number of candidate routing paths that are either minimal or non-minimal through any intermediate group.
This \textit{high path diversity} makes it challenging to select the most proper path based on a router's local information.
When proposing the Dragonfly topology, Kim et al. emphasized the importance of studying routing performance under two extreme traffics: the best scenario of UR pattern and the worst scenario of ADV+i pattern [17].
In reality, system-scale traffic patterns can be any case between these two extremes.
Furthermore, a good routing algorithm must mitigate \textit{global link congestion} as well as potential \textit{local link congestion} shown in Figure \ref{fig:adv+i}. 
Because of high path diversity and various link congestions, designing an efficient routing algorithm is complicated and challenging. 

\textbf{Routing livelock and deadlock.} 
On Dragonfly systems, when a packet traverses in the network, it relies on increasing its VC index to break the channel dependency loop.
Since the number of VCs is limited by the hardware, packets must be delivered under certain hops.
Directly applying existing RL routing (e.g., Q-routing) designed for other network topologies could lead to routing livelock and deadlock discussed in Section \ref{sec:qrtingissue}. 
An efficient MARL routing method must take these into account and deliver packets in limited hops.

\textbf{Distributed learning.}  
A key challenge in multi-agent learning is non-stationary. 
Since all agents are learning simultaneously, the environment is constantly changing from the perspective of any agent.
Hence directly applying a single agent algorithm in a multi-agent environment could lead to unstable and unsatisfactory performance. 
An efficient MARL routing must enforce certain coordination between independent agents. 

\textbf{Implementation cost.}
A good MARL routing must be lightweight in terms of both computational requirement and hardware requirement (i.e, memory consumption). 
As a result, deep reinforcement learning is unfavorable due to the heavy computation posed by neural networks, and tabular RL method is preferred. 

\section{Q-adaptive routing} \label{sec:qadp}

We present Q-adaptive routing in this section. 
It consists of three key components: (1) two-level Q-table, (2) routing with two-level Q-table, and (3) Q-table update. 
At the end, we discuss how these techniques address the challenges listed in Section \ref{sec:challenge}.

\textbf{Two-level Q-table.}
Table \ref{tab:2lqtable} depicts the newly designed two-level Q-table. 
Compared with the original Q-table (Table \ref{tab:qtable}) that is accessed according to packet destination, two-level Q-table uses both packet source and destination information.
For a Dragonfly configured with $g$ groups, $p$ nodes per $k$-radix router, each router maintains a $(g \times p) \times (k-p)$ two-level Q-table.
When routing a packet that is generated on node $n$ for group $j$, Q-values on the row of $j \times p + n$ will be evaluated for the port with the minimum estimated delivery time.
\begin{table}[ht]
\caption{Two-level Q-table}
\label{tab:2lqtable}
\includegraphics[scale=0.9]{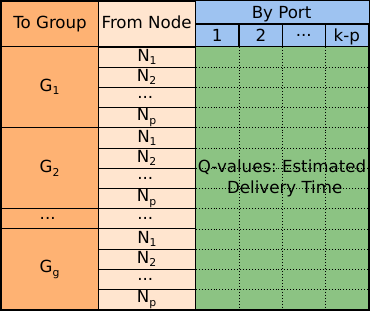}
\end{table}
Compared with the original Q-table, the two-level Q-table is smaller by $50\%$. The original Q-table contains $g \times a$ rows, whereas the two-level Q-table has $g \times p$ rows, where $a$ is the number of routers per group.
A balanced Dragonfly should be configured with $p=a/2$ \cite{08df}, thus halving the table size.
The advantages of using a smaller table are multifold:
\textit{(i)} Q-tables are stored on routers' memory, which has limited capacity.
Our two-level Q-table only requires half the memory space compared with the original Q-table, thus improves system scalability.
\textit{(ii)} The smaller table alleviates the outdated Q-value issue caused by spare updates.
In the two-level Q-table, Q-values of rarely visited destinations can also be updated along with the ones that are more frequently accessed because value update for any destination router of the same group is on the same row under the ``\verb|To Group|'' column.
\textit{(iii)} Differentiating Q-values for the same group by packet sources provides more learning information for the RL agent to take advantage of high path diversity brought by Dragonfly topology. 
% sufficient learning granularity 

\textbf{Routing with two-level Q-table.}
Figure \ref{fig:qadp-flow} presents a flow chart of the Q-adaptive routing. When a router receives a packet, it goes through the following steps to select an outbound port:
\begin{figure}[ht]
  \centering
  \includegraphics[ width=0.9\linewidth]{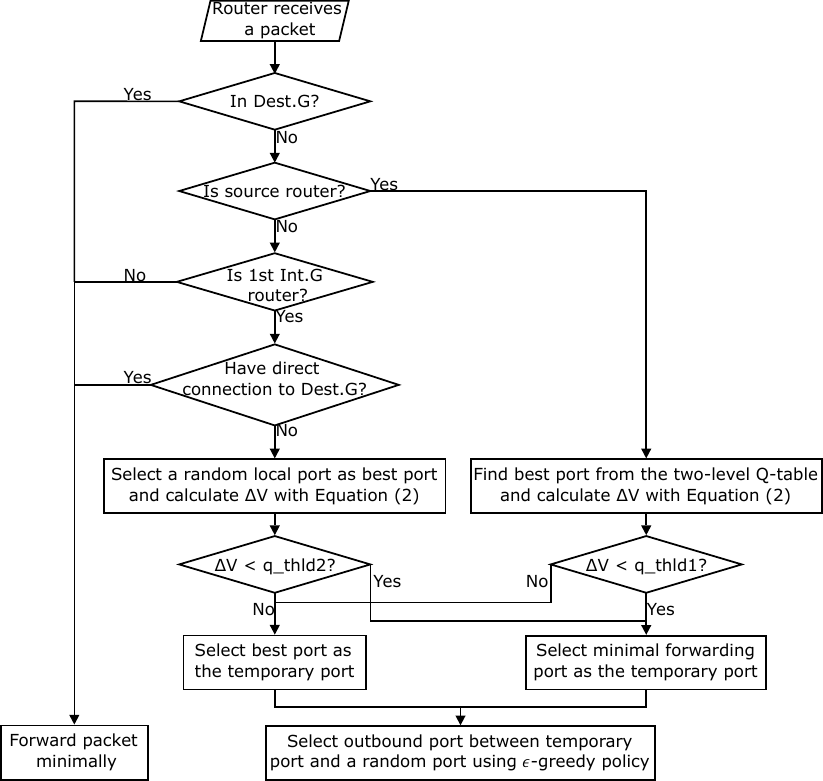}
  \caption{Flow chart for Q-adaptive routing. Dest.G and Int.G stand for destination group and intermediate group respectively.}
  \Description{}
  \label{fig:qadp-flow}
\end{figure}

\begin{enumerate}
\item If the router is in the packet's destination group, it minimally forwards the packet.

\item If the router is the packet's source router, it selects the $best\_path\_port$ with the smallest Q-value ($Q_{best\_path}$).
Next, it selects a minimal forwarding port ($min\_path\_port$) and reads the port's Q-value ($Q_{min\_path}$) from the two-level Q-table.
Then, a temporary port ($temp\_port$) is set using Equation (\ref{eq:qthld}) with \verb|threshold| $ = q\_thld1$.
Finally, the packet's outbound port is selected between $temp\_port$ and a random port using $\epsilon$-greedy policy.

\item If the router is the first intermediate group router visited by the packet, and has a direct connection to the packet's destination group, it forwards the packet minimally.
If the router is not connected to the destination group, a port for a minimal forwarding path ($min\_path\_port$) and a random local port noted as $best\_path\_port$ are selected.
Both ports' Q-values ($Q_{min\_path}$ and $Q_{best\_path}$) are read from the two-level Q-table.
Next, $temp\_port$ is set using Equation (\ref{eq:qthld}) with \verb|threshold| $ = q\_thld2$.
The final outbound port for the packet is selected between $temp\_port$ and a random port with $\epsilon$-greedy policy.

\item Otherwise, the router minimally forwards the packet towards its destination.
\end{enumerate}

\begin{equation}
\label{eq:qthld}
\begin{gathered}
\Delta V = (Q_{min\_path} -  Q_{best\_path})/Q_{min\_path} \\
temp\_port = \begin{cases}
	min\_path\_port  &\text{if $\Delta V <$} {\verb| threshold|}\\
	best\_path\_port  &\text{otherwise}
\end{cases}
\end{gathered}
\end{equation}

With the above steps, most of the routers on a routing path forward a packet minimally except at two locations: the source router and the first intermediate group router visited by a packet.
When a packet is injected into the network, the source router selects its routing path according to the two-level Q-table.
The selected path could be a minimal path or a non-minimal path that traverses an intermediate group.
When the packet is routed non-minimally and arrives at its first intermediate group router, it can be dynamically rerouted through a random router in the same group to bypass local link congestion based on the Q-values between the minimal forwarding path and the selected rerouting path.
When the routing path is chosen under the guidance of a two-level Q-table, we add a \verb|threshold| in Equation (\ref{eq:qthld}) to give a bias towards the minimal forwarding path.
Two thresholds ($q\_thld1$ and $q\_thld2$) are used for the source router and the first intermediate group router respectively. 
Both are tunable parameters.

As a result, Q-adaptive routing behaves similarly to the adaptive routing algorithms by forwarding a packet either minimally or non-minimally.
However, instead of always rerouting packets in the intermediate group as UGALn or PAR, Q-adaptive does it dynamically according to the network condition. 
Since Dragonfly is a diameter-3 topology, and only two routers on the routing path can make dynamic decisions, all packets are guaranteed to be delivered within five hops, which solves the livelock problem.
To avoid routing deadlock, Q-adaptive uses five VCs and a packet increments its VC index at every hop.

\textbf{Q-value update.}
The next challenge is how to initialize and update Q-table.
Lauer et al. proposed a \textit{distributed Q-learning} algorithm \cite{00distributedq}. 
Basically, it allows each agent to update Q-values only in a positive direction and ignores penalties from other agents. 
As such, for each router, its routing strategy is not affected by other routers' mistakes. 
While the authors proved the convergence of the distributed Q-learning algorithm in deterministic multi-agent environments, the algorithm requires initializing Q-values to very large numbers causing the downside of long convergence time. 
In \cite{07hystereticq}, Matignon et al. pointed out that totally ignoring penalties could stick the system at a sub-optimal equilibrium and hence presented \textit{hysteretic Q-learning}. 
Hysteretic Q-learning uses two learning rates to update the policy in positive and negative directions separately. 
The advantages of hysteretic Q-learning are that it does not pose any requirement for Q-value initialization and separate learning rates make the system more stable. 
Inspired by these studies, we propose to update Q-values with following equation:
\begin{equation}
\label{eq:qadp}
\begin{gathered}
\delta = r + Q_y - Q_x, x \text{ and } y \in N \\
Q_x = \begin{cases}
	Q_x + \alpha * \delta &\text{if $\delta < 0$}\\
	Q_x + \beta * \delta  &\text{otherwise} \\
\end{cases}
\end{gathered}
\end{equation}
Here, $\alpha$ and $\beta$ are the positive and negative learning rates, and $r$ is the reward, defined as the packet traveling time between neighbor routers of $x$ and $y$. 
% Q-values can be initialized as the packet delivery time without any congestion through a minimal routing path.

In summary, Q-adaptive routing is a fully distributed MARL algorithm.
Each router acts as an independent agent that learns to route by interacting with the environment using its two-level Q-table.
Not having any shared information between routers saves link bandwidth, which makes Q-adaptive routing easy to be implemented and scaled.
Meanwhile, Q-adaptive routing addresses the challenges discussed in Section \ref{sec:challenge}:

\textbf{Topology uniqueness.}
The existing adaptive routing method relies on local information to choose among up to four random routes (two minimal and two non-minimal) without exhaustively considering all possible paths \cite{crayxcnetwork}. 
In contrast, Q-adaptive routing addresses Dragonfly's high path diversity feature by using the two-level Q-table to select the best path among all viable candidates and find underestimated solutions with $\epsilon$-greedy exploration.
Moreover, the corresponding table updating techniques refresh Q-values in a timely fashion such that a router can learn to reroute packets either locally or globally to bypass network congestion.
Furthermore, learning performance can be flexibly tuned with two learning rates ($\alpha$ and $\beta$), and a proper bias toward minimal path can be controlled with two thresholds ($q\_thld1$ and $q\_thld2$).

\textbf{Routing livelock and deadlock.}
Q-adaptive routing guarantees a packet to be delivered efficiently within five hops.
Therefore, routing livelock never happens.
Limited routing path length also makes the system deadlock free with five VCs.

\textbf{Distributed learning.}
While each router is an independent agent, Q-adaptive routing coordinates them through two-level Q-table updating with two learning rates. 
By leveraging hysteretic Q-learning, Q-adaptive routing guarantees a fast learning convergence mainly controlled by $\alpha$ and maintains the converged state stable with $\beta$.

\textbf{Implementation cost.}
Q-adaptive routing requires no specific hardware implementations nor heavy computational resources.
The proposed two-level Q-table halves the original Q-table size. 
The small-sized Q-table improves the algorithm's scalability, as well as alleviates the outdated Q-value problems inherent in the original Q-routing algorithm.
Q-adaptive routing can be easily deployed on current high-radix Dragonfly routers by replacing the routing table with the proposed two-level Q-table of size $O(g*p*k)$ \cite{isca06blackwidow}.
The two-level Q-table updating only involves basic arithmetic operations and the reward transferred between neighbor routers could be embedded into the credit-based flow control flits or other network control packets.  

\section{Evaluation} \label{sec:1k-res}

We use and expand the open-source, community-built simulation toolkit SST/Merlin \cite{sst}. 
SST/Merlin is a high-fidelity network modeling toolkit supporting several interconnect topologies including Dragonfly. 
It provides a detailed flit-level input-output queued high-radix router model with credit-based flow control. 
In this work, we enhance the toolkit by implementing different traffic patterns. 
We implement our Q-adaptive routing, PAR, UGALg, and UGALn in the toolkit. 
In our evaluation, we compare Q-adaptive with five routing methods: minimal routing (MIN), VALn non-minimal routing, UGALg, UGALn, and PAR.
Note that MIN and VALn represent optimal routing strategies under UR and ADV+i traffic patterns respectively.
UGAL and PAR are the existing adaptive routing mechanisms deployed on production Dragonfly systems and serve as the baseline for comparison study.

We evaluate Q-adaptive routing on two systems: 1,056-node and 2,550-node Dragonfly systems. 
In this section, we examine the results achieved on the 1,056-node system. 
In Section \ref{sec:2k-res}, we will present performance results on the scale-up 2,550-node case to demonstrate the scalability and feasibility of the Q-adaptive routing. 

\subsection{Experimental Setup}

Each system is configured using 128B single flit packets to avoid the effect of flow control on routing results with single packet messages.
Routers are configured with VC buffer size of 20 packets, and all the link bandwidth is 4GB/s with local link latency of 30ns \cite{crayxcnetwork}, and global link latency of 300ns to keep the 1:10 ratio as mentioned in previous works \cite{08df}\cite{isca09indirectadp}\cite{dffarendcongest}.
All adaptive routing algorithms are set with zero bias towards minimal routing and use local output queue occupancy plus the used credit count to estimate routing path congestion.

For the 1,056-node system, all results are evaluated with  $\alpha = 0.2$, $\beta=0.04$, $\epsilon=0.001$, $q\_thld1=0.2$, $q\_thld2=0.35$ in this section.
The values are selected to balance the trade-off between convergence speed and network peak performance. 
Q-values are initialized to the theoretical packet delivery time without any congestion through a minimal routing path.

In our experiments, different routing methods are compared under various traffic patterns. 
For each pattern, messages are generated at an adjustable interval to create different network conditions, ranging from a lightly loaded system to a fully loaded system. 
We simulate network usage according to \emph{offered load}, which is defined as the ratio between the message generation rate and the system-wide total injection bandwidth.
Thus, offered load is between 0 and 1, where 0 means an empty network and 1 means a fully loaded system.

\begin{figure*}[ht]
  \centering
  \includegraphics[scale=0.9]{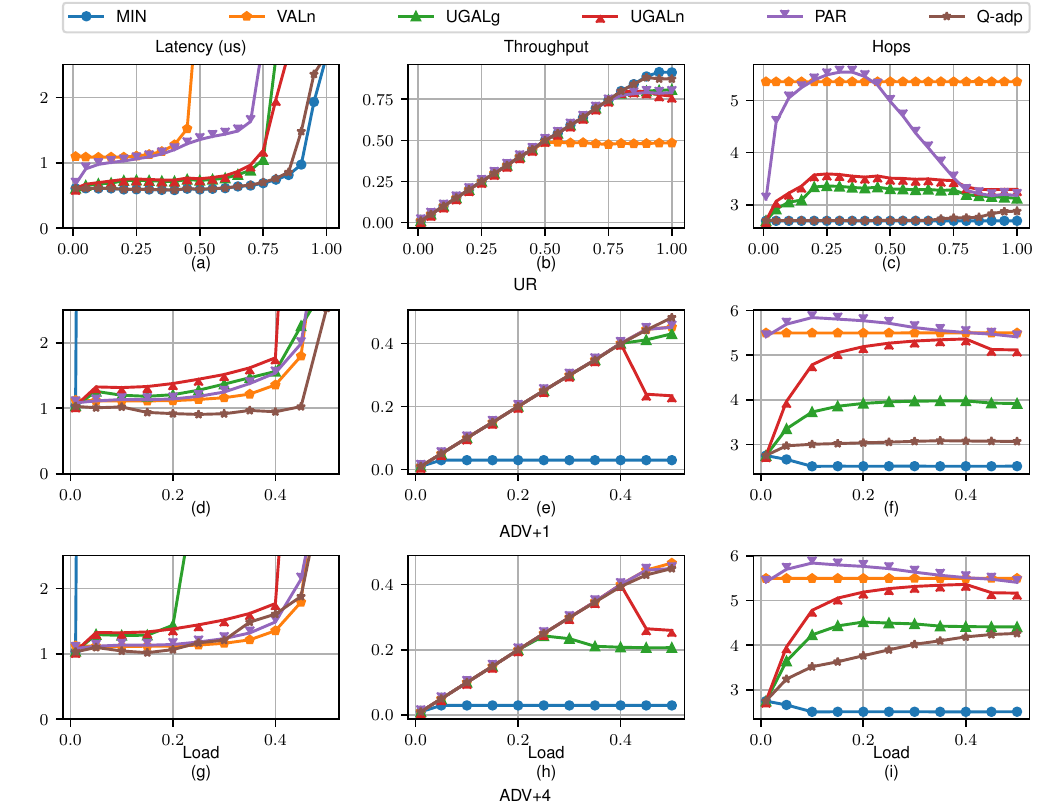}
  \caption{Q-adaptive on the 1056-node Dragonfly}
  \Description{}
  \label{fig:qadp-load}
\end{figure*}

\subsection{Evaluation metric}

We use two evaluation metrics for performance comparison: (1) \emph{packet latency}, which denotes the packet traveling time from its generation to delivery; and (2) \emph{system throughput}, which is measured as the aggregated message receiving rate across the system.
Therefore, system throughput is always smaller than the \textit{offered load}, with 0 means all packets are congested or blocked on the network.

% For the rest of the section, we present Q-adaptive routing results by analyzing its performance under different loads (Sec. \ref{sec:qadploads}), examining tail latency (Sec \ref{sec:1ktail}) , discussing system convergence (Sec. \ref{sec:qadpconverge}), and exploring its transient performance under a dynamically changing system (Sec. \ref{sec:qadpdynamic}).
% \textcolor{blue}{and demonstrating its scalability with scale-up performance analysis (Sec. \ref{sec:2ksystem})}.

\subsection{Routing under Different Loads} \label{sec:qadploads}

Figure \ref{fig:qadp-load} depicts Q-adaptive routing results for the UR and ADV+1, ADV+4 traffic patterns.
For each traffic pattern, three plots show the packet latency, system throughput, and packet hop count under different offered loads.
The results shown in Figure \ref{fig:qadp-load} are collected as the arithmetic average over \SI{100}{\micro\second} after the system becomes stable.
Adaptive and minimal, non-minimal routing results are also plotted for comparison.

\textbf{UR pattern:} 
Under the UR traffic pattern, Q-adaptive outperforms all the adaptive routing algorithms regarding system throughput shown in Figure \ref{fig:qadp-load}(b).
Under the maximum load, Q-adaptive reaches 88.25\% system throughput, which is 6.60\%, 10.51\%, and 8.32\% higher than UGALg, UGALn, and PAR respectively.
Meanwhile, system throughput of Q-adaptive is only 3.29\% smaller than that of the optimal minimal routing.

Q-adaptive also outperforms adaptive routing algorithms regarding packet latency.
In Figure \ref{fig:qadp-load}(a), the packet latency of Q-adaptive routing is always smaller than that of adaptive routing methods.
All the adaptive routing algorithms show obvious network congestion starting from the offered load of 0.8 with a rapid packet latency increase.
On the other hand, Q-adaptive routing can handle this load with an average packet latency of \SI{0.76}{\micro\second}, which is 3.43x, 2.59x, and 5.22x smaller than that of UGALg, UGALn, and PAR respectively.
Furthermore, the average latency of Q-adaptive is close to the optimal minimal routing's \SI{0.74}{\micro\second} under the load of 0.8.

\begin{figure*}[ht]
  \centering
  \includegraphics[scale=0.9]{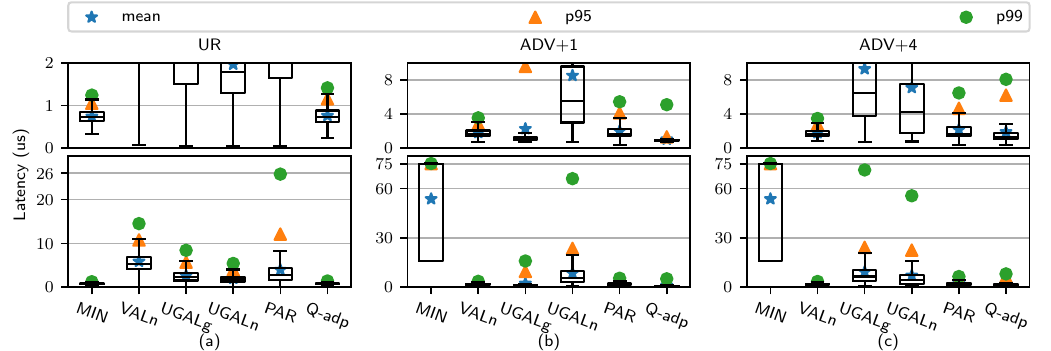}
  \caption{Packet latency distribution on the 1,056-node Dragonfly with zoom-in plot shown on top. The central box presents the latency range from $Q1$ (first quartile) to $Q3$ (third quartile) with the median shown as a black line. 
  The upper and lower whiskers outside of the central box are $1.5\times IQR$ above and below $Q3$ and $Q1$ respectively. Average packet latency and 95th, 99th percentile latency are also plotted.
}
  \Description{}
  \label{fig:qadp-1k-sys-pattern-dist}
\end{figure*}

\textbf{ADV+1 pattern:}
Figure \ref{fig:qadp-load}(e) shows that Q-adaptive routing can achieve a maximum of 48.20\% system throughput, which is 5.15\%, 8.20\%, and 3.09\% higher than UGALg, UGALn, and PAR.
Noticeably, under the maximum load, Q-adaptive routing even improves system throughput by 3.0\% compared with the optimal VALn non-minimal routing.
Q-adaptive outperforms VALn routing because it only forwards packets non-minimally in intermediate groups when it is necessary.
Since ADV+1 introduces the least local link congestion, most of the packets should be minimally forwarded in intermediate groups, whereas VALn always reroutes packets through a random router causing network bandwidth waste.
As a result, Q-adaptive uses the network more efficiently by sending packets with an average of 3.06 hops shown in Figure \ref{fig:qadp-load}(f), which is 1.80x smaller than that of VALn routing.

Besides achieving the highest throughput, using shorter routing paths makes Q-adaptive routing deliver packets in the shortest time.
In Figure \ref{fig:qadp-load}(d), all routing methods except Q-adaptive show network congestion at the offered load of 0.45 with a sharp average packet latency increase.
Under this load, the average packet latency of Q-adaptive routing is only
\SI{1.03}{\micro\second}. 
% which is 2.20x, 8.33x, 1.94x, and 1.75x smaller than that of UGALg, UGALn, PAR, and VALn routing.

% Therefore, Q-adaptive can learn to route non-minimally under the ADV+1 traffic pattern.
% Meanwhile, Q-adaptive does not waste link bandwidth in the intermediate groups and achieves the best performance.

\textbf{ADV+4 pattern:}
The ADV+4 traffic pattern introduces the heaviest local link congestion.
Figure \ref{fig:qadp-load}(i) shows that Q-adaptive routes packets with an average of 4.27 hops at the offered load of 0.5, which is larger than the 3.06 hops under the ADV+1 traffic pattern.
This shows that Q-adaptive can reroute packets in intermediate groups with an additional hop to bypass local link congestion. 
As a result, Q-adaptive achieves up to 44.93\% system throughput, which is only 1.69\% less than the optimal VALn non-minimal routing.
Q-adaptive and PAR have similar average packet latency performance shown in Figure \ref{fig:qadp-load}(g).
Both Q-adaptive and PAR routing deliver packets using less time than UGALg and UGALn. 

To summarize, Q-adaptive outperforms adaptive routing algorithms with near-optimal performance.
It learns to route minimally when traffic is uniform and to use intermediate groups and routers when the traffic pattern is adversarial. 

% In conclusion, the evaluation of the UR and ADV+1, ADV+4 traffic patterns show that Q-adaptive routing is capable of choosing the best routing strategy according to traffic pattern and system load.
% As a result, Q-adaptive can outperform existing adaptive routing methods and achieve near-optimal performance.

\subsection{Tail Latency} \label{sec:1ktail}

Figure \ref{fig:qadp-1k-sys-pattern-dist} shows the packet latency distribution when the load is set to 0.8 under UR and 0.45 under ADV+i. 
These plots show the average packet latency with the 95th and 99th percentile latency. 
% Because underutilized network (load close to 0) introduces few challenges to the routing algorithms and over-utilized network (load close to 1) leads to network congestion with large packet latency making comparison less meaningful, Figure \ref{fig:qadp-1k-sys-pattern-dist} depicts cases of the UR traffic pattern under the load of 0.8 and the ADV+1, ADV+4 traffic patterns under the load of 0.45.

\textbf{UR pattern:}
In Figure \ref{fig:qadp-1k-sys-pattern-dist}(a), Q-adaptive has the smallest 99th percentile latency of \SI{1.42}{\micro\second}, which is 5.92x, 3.82x, and 18.18x smaller than that of UGALg, UGALn, and PAR respectively.
Further, this 99th percentile latency is very close to the theoretic optimum, i.e., \SI{1.25}{\micro\second} achieved by the minimal routing. 

\textbf{ADV+1 pattern:}
In Figure \ref{fig:qadp-1k-sys-pattern-dist}(b), the 99th percentile latency of Q-adaptive routing is \SI{5.10}{\micro\second}, which is 3.12x, and 12.95x smaller than that of UGALg and UGALn respectively.
UGALn performs poorly because the ADV+1 pattern introduces the least local link congestion.
Under this pattern, packet rerouting in intermediate groups performed by UGALn wastes local link bandwidth and delays packet delivery.
PAR achieves the 99th percentile latency of \SI{5.44}{\micro\second}, which is close to the \SI{5.10}{\micro\second} obtained by Q-adaptive. 
However, the latency distribution achieved by Q-adaptive is much more compact than that of PAR. 
In other words, Q-adaptive outperforms PAR and UGAL in terms of the overall latency distribution and tail latency.

\textbf{ADV+4 pattern:}
In Figure \ref{fig:qadp-1k-sys-pattern-dist}(c), the 99th percentile latency of Q-adaptive is \SI{8.08}{\micro\second}, which is 8.83x and 6.89x smaller than that of UGALg and UGALn.
While the 95th and 99th percentile latencies achieved by PAR are slightly less than those obtained by Q-adaptive, we shall point out that Q-adaptive results in a more concentrated packet latency distribution. 
When using Q-adaptive, 80.99\% of packets have a latency of less than \SI{2}{\micro\second}, whereas only 63.69\% of packets have such a low latency when using PAR. 

Put together, when comparing different routing algorithms under UR and ADV+i patterns, Q-adaptive outperforms UGALg, UGALn, and PAR in terms of average latency and tail latency distribution. 

\subsection{Convergence} \label{sec:qadpconverge}

\begin{figure}[ht]
  \centering
  \includegraphics[]{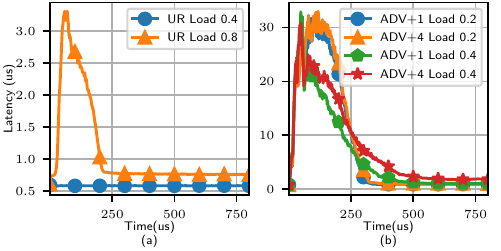}
  \caption{Convergence of Q-adaptive. Q-adaptive routing can converge under  \SI{500}{\micro\second} under different traffic patterns and loads. }
  \Description{}
  \label{fig:qadp-converge}
\end{figure}

Figure \ref{fig:qadp-converge} demonstrates how the packet latency evolves with a system starting from an empty network. 

For the UR traffic pattern in Figure \ref{fig:qadp-converge}(a), Q-adaptive routing can nearly handle the lower offered load case directly with a negligible learning period.
For the higher offered load case, network traffic generated at the system start time (\SI{0}{\micro\second}) can quickly congest the system and leads to a dramatic packet latency surge.
However, routers are capable of learning an optimal routing policy within approximately \SI{200}{\micro\second} to bring the system to a stable state with an average packet latency of  \SI{0.75}{\micro\second}.
Compared with the UR pattern, ADV+1 and ADV+4 patterns require longer convergence time for routers to learn to route non-minimally as shown in Figure \ref{fig:qadp-converge}(b).
Furthermore, compared with the ADV+1 traffic pattern, ADV+4 requires routers to spend some additional time to learn to alleviate local link congestion through packet rerouting in intermediate groups.
Nevertheless, Q-adaptive routing can still find its optimal equilibrium for both traffic patterns under different loads within \SI{500}{\micro\second}.

In summary, Q-adaptive routing can converge within a small amount of time under both UR and ADV+i traffic patterns.

\subsection{Dynamic Loads} \label{sec:qadpdynamic}

\begin{figure}[ht]
  \centering
  \includegraphics[]{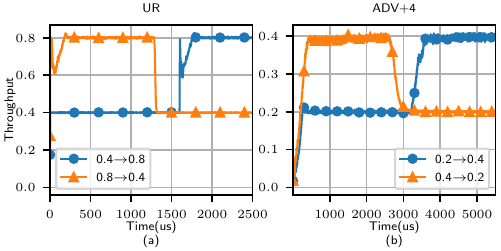}
  \caption{Q-adaptive under varying loads}
  \Description{}
  \label{fig:qadp-dynamicload}
\end{figure}

Figure \ref{fig:qadp-dynamicload} depicts how system throughput evolves starting with an empty network and a traffic pattern with varying offered load. 

In Figure \ref{fig:qadp-dynamicload}(a), the UR traffic pattern offered load is increased from 0.4 to 0.8 at \SI{1600}{\micro\second}, and Q-adaptive needs to go through a learning period for about \SI{156}{\micro\second} to gradually adapt to the higher load.
Note that the \SI{156}{\micro\second} learning period is shorter than the \SI{200}{\micro\second} learning period when the system starts from an empty network shown in Figure \ref{fig:qadp-converge}(a).
When the UR offered load is decreased from 0.8 to 0.4 at \SI{1280}{\micro\second}, the system instantly responds to the change. 
In Figure \ref{fig:qadp-dynamicload}(b), when the ADV+4 traffic pattern increases its offered load from 0.2 to 0.4 at \SI{3215}{\micro\second}, \SI{455}{\micro\second} is required to let Q-adaptive adjust to this change.
Similarly, \SI{440}{\micro\second} is required when the pattern decreases its load from 0.4 to 0.2 at \SI{2610}{\micro\second}.
Both cases require less time than the \SI{500}{\micro\second} learning period shown in  Figure \ref{fig:qadp-converge}(b).

\section{Case Study on 2,550-node system} \label{sec:2k-res}

\begin{figure*}[ht]
  \centering
  \includegraphics[scale=0.9]{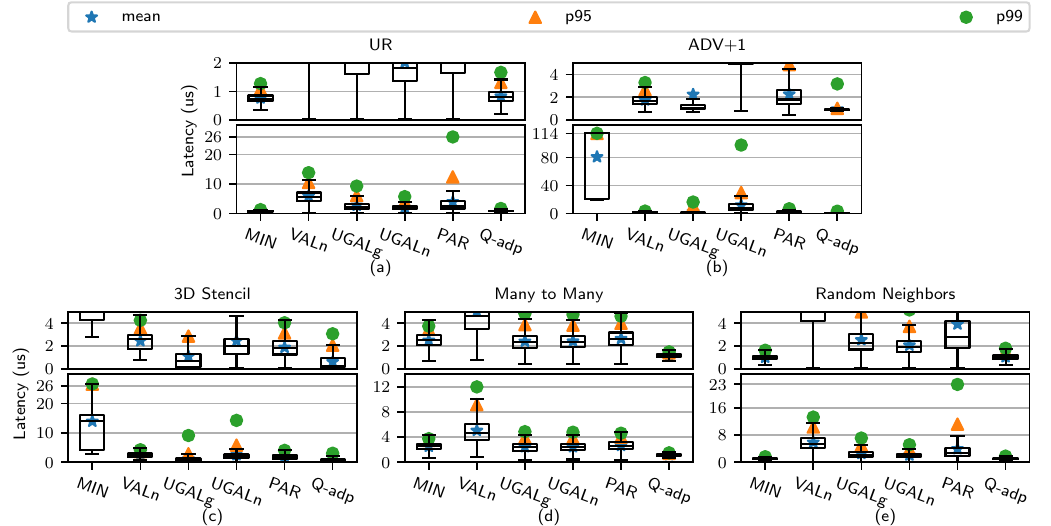}
  \caption{Packet latency distribution on the 2,056-node Dragonfly  with zoom-in plot shown on top.}
  \Description{}
  \label{fig:qadp-2k-sys}
\end{figure*}

Now we present a scale-up study of Q-adaptive routing on a 2,550-node Dragonfly system (configuration shown in Table \ref{tab:1056DF}). 
In this case study, we evaluate Q-adaptive routing under both system-wide extreme traffic patterns (UR and ADV+1) and three common HPC communication patterns: 

\begin{itemize}
    \item 3D Stencil: stencil computation is an important class of algorithms in scientific computing and 3D Stencil is a representative one-to-many pattern \cite{stencil2}\cite{stencil1}. 
    Under this pattern, the system is arranged in a 5$\times$10$\times$51 3D grid and each node communicates with its six neighbors along three dimensions.
    
    \item Many to Many: this pattern is representative in applications that perform parallel Fast Fourier Transforms such as pF3D \cite{pf3d}, NAMD \cite{namd}, and VASP \cite{vasp}.
    Under this pattern, the system is arranged in a  5$\times$10$\times$51 3D grid. 
    Nodes along Z-axis are grouped into the same communicator of 51 nodes performing \textbf{\textit{all-to-all}} operations.

    \item Random Neighbors: the pattern mimics computation-aware load-balancing operation performed by HPC applications such as NAMD \cite{namd}.
    Under this pattern, each node uniformly spreads communications with 6--20 randomly selected targets.

\end{itemize}

In this section, all the patterns are evaluated with the same hyperparameter values of $\alpha = 0.2$, $\beta=0.04$, $\epsilon=0.001$, $q\_thld1=0.05$, $q\_thld2=0.4$.
Figure \ref{fig:qadp-2k-sys} shows the packet latency distribution under different traffic patterns. 

\textbf{UR pattern:}
As shown in Figure \ref{fig:qadp-2k-sys}(a), the average packet latency of Q-adaptive routing is \SI{0.84}{\micro\second}, which is 3.24x, 2.40x, 4.67x smaller than that of UGALg, UGALn, and PAR respectively.
The 99th percentile latency of Q-adaptive routing is \SI{1.67}{\micro\second}, which is significantly smaller than the adaptive routing methods.
Additionally, Q-adaptive has the near-optimal performance compared with the optimal minimal routing, whose average packet latency and 99th percentile latency are \SI{0.77}{\micro\second} and \SI{1.28}{\micro\second}.

\textbf{ADV+1 pattern:}
In Figure \ref{fig:qadp-2k-sys}(b), the 
average packet latency of Q-adaptive is \SI{0.96}{\micro\second}, which is 2.33x, 12.35x, 2.35x smaller than that of UGALg, UGALn, and PAR respectively.
Q-adaptive also has the smallest 99th percentile latency of \SI{3.17}{\micro\second} among all other routing methods.
As a result, same as the 1,056-node system case, Q-adaptive outperforms the optimal VALn non-minimal routing, whose average packet latency is \SI{1.75}{\micro\second} and 99th percentile latency is \SI{3.29}{\micro\second}.

\textbf{3D Stencil pattern:}
Figure \ref{fig:qadp-2k-sys}(c) shows that Q-adaptive has the smallest average packet latency of \SI{0.62}{\micro\second}, which is 1.77x smaller than the second-best UGALg's average latency.
Additionally, the 99th percentile latency of Q-adaptive, \SI{3.08}{\micro\second}, is also the smallest among all routing algorithms and is 1.31x smaller than the second-best PAR's 99th percentile latency.

\textbf{Many to Many pattern:}
Figure \ref{fig:qadp-2k-sys}(d) shows that Q-adaptive has the smallest average packet latency of \SI{1.15}{\micro\second}, which is 2.10x smaller than the second-best UGALg's average packet latency.
The 99th percentile latency of Q-adaptive is \SI{1.50}{\micro\second}, which is 2.50x smaller than the second-best minimal routing's 99th percentile latency. 
As a result, Q-adaptive outperforms all other routing algorithms regarding both average and 99th percentile packet latency.

\textbf{Random Neighbors pattern:}
Under this pattern, the traffic is uniformly spread across the network.
Thus minimal routing is the optimal routing solution with the smallest average packet latency of \SI{1.01}{\micro\second} and the smallest 99th percentile latency of \SI{1.64}{\micro\second} shown in Figure \ref{fig:qadp-2k-sys}(e).
However, Q-adaptive offers near-optimal performance with \SI{1.04}{\micro\second} average packet latency and  \SI{1.81}{\micro\second} 99th percentile latency.
Although UGALn performs best among adaptive routing algorithms, its average and 99th percentile latency are 1.99x and 2.86x larger than that of Q-adaptive routing.

Overall, we make two key observations from this scale-up case study. First, Q-adaptive routing is capable of achieving better performance than existing dragonfly routing algorithms under system-wide extreme traffic patterns and three representative HPC communication patterns. 
Second, Q-adaptive is capable of scaling on larger systems. 

\section{Conclusion} \label{sec:conclude}

Dragonfly is a promising high-radix interconnect topology for large-scale HPC systems.
The high path diversity and potential link congestion inherent in Dragonfly pose significant challenges when routing packets among many candidate paths. 
Existing adaptive routing methods use local router information (i.e., local queue occupancy) 
% to deduce global path congestion,
and select a path among a limited number of paths, which can lead to inefficient link usage and low system throughput. 
In this work, we have presented Q-adaptive routing, a multi-agent reinforcement learning based scheme, for dragonfly systems. 
The proposed Q-adaptive routing features by its small yet effective two-level Q-table design, along with its value updating and routing mechanisms.
These techniques enable Q-adaptive routing to effectively learn to choose a proper path among all the viable paths for packet routing on Dragonfly. 
We have evaluated Q-adaptive routing on both 1k and 2k systems through extensive simulations.
Our results indicate that Q-adaptive outperforms existing adaptive routing methods by up to 10.5\% in network throughput and more than 5x reduction in packet latency. 
Moreover, Q-adaptive routing achieves better tail latency than the existing adaptive routing methods. 

In this study, we have focused on analyzing Q-adaptive routing under system-wide traffic patterns. 
Given the promising results achieved in this study, part of our future work is to analyze Q-adaptive on individual applications and further to investigate inter-job interference by using Q-adaptive routing. 

%%
%% The acknowledgments section is defined using the "acks" environment
%% (and NOT an unnumbered section). This ensures the proper
%% identification of the section in the article metadata, and the
%% consistent spelling of the heading.
\begin{acks}
We thank our shepherd and the HPDC reviewers for their valuable feedback and insightful comments. 
This work is supported in part by US National Science Foundation grants CNS-1717763, CCF-1618776.
\end{acks}

%%
%% The next two lines define the bibliography style to be used, and
%% the bibliography file.
\bibliographystyle{ACM-Reference-Format}
\bibliography{sample-base}

%%
%% If your work has an appendix, this is the place to put it.
\appendix

\end{document}